\documentclass[traditabstract]{aa}
\usepackage[dvips]{graphicx}

\usepackage{rotating}
\usepackage{longtable}
\usepackage{amssymb,amsmath}
\newcommand{\be}{\begin{equation}}
\newcommand{\ee}{\end{equation}}
\usepackage{latexsym}
\usepackage{verbatim}
\newcommand{\simless}{\mathbin{\lower 3pt\hbox {$\rlap{\raise 5pt\hbox{$\char'074$}}\mathchar"7218$}}} %< or of order
\newcommand{\simgreat}{\mathbin{\lower 3pt\hbox
     {$\rlap{\raise 5pt\hbox{$\char'076$}}\mathchar"7218$}}} %> or of order

\newcommand{\cmq}{cm$^{-3}$}
\newcommand{\um}{$\mu m$}
\newcommand{\sori}{$\sigma$~Ori}
\newcommand{\Msun}{M$_\odot$}
\newcommand{\Lsun}{L$_\odot$}
\newcommand{\Ha}{H$\alpha$}
\newcommand{\Hb}{H$\beta$}
\newcommand{\Hg}{H$\gamma$}
\newcommand{\Hd}{H$\delta$}
\newcommand{\Pab}{Pa$\beta$}
\newcommand{\Pag}{Pa$\gamma$}
\newcommand{\Brg}{Br$\gamma$}

\newcommand{\Lacc}{$L_{acc}$}
\newcommand{\Macc}{$\dot M_{acc}$}

\newcommand{\Myr}{M$_\odot$/y}

\begin{document}

\title{ X-shooter observations of  the accreting brown dwarf  J053825.4-024241 
\thanks{
Based on observations collected at the European Southern Observatory, Chile.
Program 084.C-0269(A)}
}

\author{
E. Rigliaco\inst{1,2},
A. Natta\inst{1,3},
S. Randich \inst{1},
L. Testi\inst{1,4},
E. Covino\inst{5},
G. Herczeg\inst{6},
\and
J. M. Alcal\'a\inst{5}
}

\institute{
    Osservatorio Astrofisico di Arcetri, INAF, Largo E. Fermi 5,
    50125 Firenze, Italy
\and
Universit\`a  di Firenze, Dipartimento di Astronomia, Largo E.Fermi 2,
    I-50125 Firenze, Italy
\and
School of Cosmic Physics, Dublin Institute for Advanced Studies, Dublin 2, Republic of Ireland
\and
ESO, Karl-Schwarschild Strasse 2, 85748 Garching bei M\"unchen, Germany
\and
Osservatorio Astronomico di Capodimonte, INAF, 80131 Napoli, Italy
\and
Max-Planck-Institut fur extraterrestriche Physik, 85741 Garching bei M\"unchen, Germany.
}

\offprints{erigliaco@arcetri.astro.it}
\date{Received 2010 October 27; accepted 2010 December 09}

\authorrunning{Rigliaco et al.}
\titlerunning{X-shooter observation of J053825.4-024241}

\abstract
{We present the first observations of a probable brown dwarf, 
obtained with the new spectrograph X-shooter mounted 
on the UT2@VLT. The target (2MASS J053825.4-024241) is a 0.06 \Msun\ object
in the star-formation region \sori. The X-shooter spectrum covers
simultaneously the whole range from UV to NIR (300-2500 nm). 
The J053825.4-024241 spectrum
is rich in emission lines that are typical of accreting young object and clearly shows
the Balmer jump.
Moreover, many photospheric atomic and molecular absorption lines 
yield the spectral type and confirm that the object is young.
We compute the mass accretion rate from all available observed accretion diagnostics.
We find that 
there is a large spread in the \Macc\ values (up to a factor 40) 
that is not caused by variability; 
some of this spread
may be intrinsic, i.e., owing to different 
physical conditions of the emitting region for the same \Macc. 
However, within the large error bars all \Macc\ measurements agree, 
and the mean value is log\Macc $\sim -9.86\pm 0.45$ \Myr.
The hydrogen Balmer lines are clearly detected up to $n=25$. Their ratios
suggest that the emitting region is cold (T$\sim 2000-3000$ K), 
dense and in thermal equilibrium (LTE), and that the lines are optically thick up to $n\sim 21$.
We briefly discuss the implications of this result for magnetospheric accretion models.

}


\maketitle

\section {Introduction}

Accretion and ejection of matter play a fundamental role in shaping the
structure and evolution of proto-planetary disks. In the last
years, our understanding of
these processes has progressed significantly, confirming the role played by
accretion  and magnetic fields, but at the same
time raising new questions. A large number of observational diagnostics need to be
observed simultaneously to avoid the problems related to
the time variability that characterize
pre-main sequence stars.
X-shooter, the new spectrometer at VLT/ESO, is the optimum instrument
for this purpose, with its large wavelength range (300-2500 nm) that is
covered simultaneously.

This letter reports  the first observations of a brown dwarf carried out with X-shooter. 
The target J053825.4-024241 (J0538 in the following) in the \sori\ star-forming region 
(D$\sim$360 pc, B\'ejar et al.~2001),
was first identified as a photometric sub-stellar candidate 
by B\'ejar et al.~(2004), and was then extensively 
studied by Caballero et al.~(2006), who concluded that
J0538 is a likely cluster member, with heliocentric 
radial velocity $v_h = 32 \pm 13 \, \rm{km \, s^{-1}}$,
spectral type M6$\pm 1$, and  a mass 
estimated from evolutionary models of M$\sim$0.06\Msun.  
J0538  is detected by {\it{Spitzer}} at all IRAC wavelengths and at 24 \um\  
with MIPS (Hern\'andez et al.~2007). 
Its spectral energy distribution (SED) shows a clear excess 
over the photospheric emission from the K band, which is indicative of a circumstellar disk 
sorrounding the brown dwarf.

\section {Observations and analysis}

The observations were obtained with X-shooter within the INAF/GTO program 
 on star-forming regions. %(P.I. J.M.Alcal\`a).
Observations were performed in visitor mode during the night of the 2009 December 23; 
exposure time was 6$\times$900 sec, which allows us to reach a signal-to-noise ratio 
in the continuum 
from $\sim$ 7 (UVB-arm) to $\sim$ 30 
(VIS-arm) and $\sim$ 80  (NIR-arm),
depending on the echelle order.

\begin{figure} [h!]
   \centering
%   \resizebox{\hsize}{!}{\includegraphics[]{FIGpaper_LieK.epsi}}
	\includegraphics[width=0.4\textwidth]{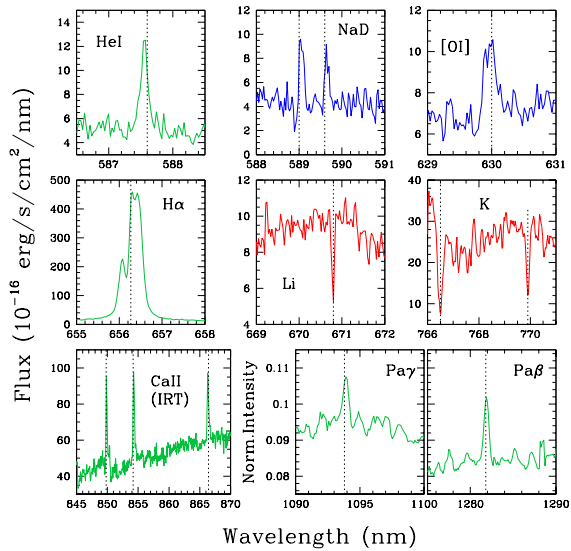}
           \caption{\small Spectra of selected lines, as labelled.
Green spectra show lines used to derive the mass accretion rates, 
red spectra show absorption features, and blue spectra wind diagnostic lines. 
%Color versions of the Figures are available on the online journal.
}
\label {spectra}
\end{figure}

\begin{figure}[!htb]
   \centering
%   \resizebox{\hsize}{!}{\includegraphics[]{bj.epsi}}
   \includegraphics[width=0.45\textwidth]{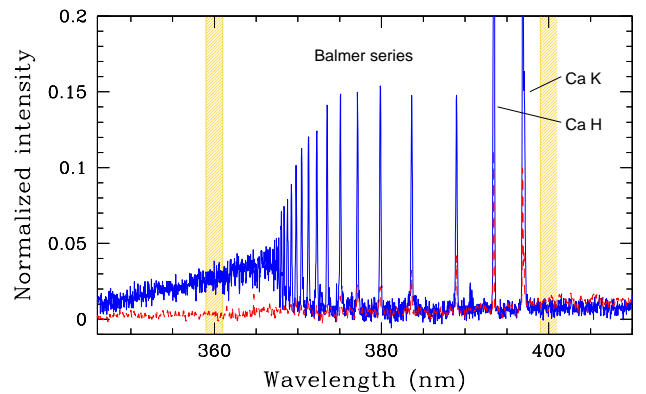}
        \caption{\small Spectrum of J0538 (solid/blue line) in the region near the 
        Balmer jump. 
        Dashed/red line refers to a template spectrum of a class III low-mass 
        star  in \sori\ with 
        spectral type M6.5. 
        The two shaded intervals indicate where the fluxes 
        @ 360 nm and  @ 400 nm 
        have been derived. 
        %A color version of the Figure is available on the online journal.
        }
\label {jump}
\end{figure}

The target was observed with the $11''\times1.0''$ slit in the UVB-arm 
and with $11''\times0.9''$ slit in the VIS and NIR-arms. 
The spectral resolution 
R is $\sim$ 5100 over the UVB and the NIR-arm, 
and $\sim$ 8800
in the VIS-arm.
Raw data were reduced using the EsoRex pipeline (version 0.9.4) following the standard 
steps, which include bias subtraction, flat-fielding, optimal extraction, wavelength 
calibration, sky subtraction, and flux-calibration. \\
The extraction of the 1D spectra and the subsequent data analysis was performed 
with the IRAF \footnote{{\sc iraf} is distributed by National Optical Astronomy
Observatories, which is operated by the Association of Universities
for Research in Astronomy, Inc., under cooperative agreement with the
National Science Foundation.} package.
The flux calibration of the star was performed using a spectrophotometric standard 
(GD71-V13.06)
observed close in time and airmass to J0538. 
The spectrum of J0538 shows strong emission lines of H, \ion{He}{i}, 
\ion{Ca}{ii}, [\ion{O}{i}] 
as well as photospheric  atomic absorption lines of Li and K
and broad molecular absorption bands such as TiO. Balmer continuum
emission is also clearly detected.
Some regions of the stellar spectrum in the three arms 
are plotted in Figs.~\ref{spectra} and \ref{jump}. 
The properties of selected lines are given in Table~\ref{parameters}.

\section {The brown dwarf J053825.4-024241}

J0538 (SO500 in Hern\'andez et al.~2007)  
shows spectroscopic features that are typical of low-gravity atmospheres 
(rather weak \ion{K}{i}$\lambda\lambda$766.5,769.9 nm and \ion{Na}{i}$\lambda\lambda$818.3,819.5 nm absorption lines), which support the assumption of the youth of the object 
and its membership to the \sori\ cluster. 
Lithium at $\lambda$670.8 nm 
is detected in absorption, as expected.
However, the observed lithium pseudo-equivalent width (pEW) ($\sim$0.33$\pm$0.02\AA) 
is found to be rather low with respect to 
the theoretical pEWs obtained by Palla et al.~(2007) for undepleted lithium. 
This discrepancy possibly stems from a lack of understanding 
of the details of the formation of the lithium line, or 
of the internal structure of young low-mass stars (Johnas et al.~2007).   

The spectral type, determined 
by the comparison between the optical spectrum of the object with field dwarf template spectra 
(Kirkpatrick et al. 1993),
is M7 $\pm$ 0.5 ($\rm{T_{eff}} \sim$ 2900 K, following
the temperature scale of Luhman 1999) and agree well with 
the determination obtained by Caballero et al.~(2006). 
The optical extinction to the cluster is negligible (Oliveira et al.~2004) 
and the estimated stellar luminosity for the
adopted distance is $\sim$ 0.025 $\pm$ 0.005~\Lsun, based on the 
spectral type and on the average I and J magnitudes 
(Rigliaco et al.~2010 and references
therein).
Based on its location on the HR diagram and on a comparison with the 
Baraffe et al. (1998) evolutionary tracks, J0538 has a mass of 
about 0.06~\Msun\ (not larger than 0.08~\Msun\ considering all the uncertainties), 
and an age $\sim$1 Myr, 
apparently slightly younger than the bulk of the \sori\ stars
(e.g., Rigliaco et al.~2010).

\section {Mass accretion rate}

The X-shooter spectrum of J0538 is rich in emission in lines and continuum,
which are powered by processes related to the accretion from the
circumstellar disk onto the central object. These diagnostics
are used to derive 
the mass accretion rate 
from empirical relations between the observed line or continuum excess 
luminosity and 
the accretion luminosity or the mass accretion rates 
(see, e.g., Herczeg \& Hillenbrand~2008, hereafter HH08, and references therein).
These relations were calibrated on a relatively small sample
of well-studied objects, mostly T Tauri stars, for
which the accretion luminosity was derived by fitting the observed optical veiling 
with magnetospheric accretion models
(Gullbring et al.~1998; Calvet et al.~2004),
or, for brown dwarfs, the \Ha\ profiles (Muzerolle et al.~2005).

We compute the \Macc\ values from the \Ha\ luminosity according to the
relation   in Fang et al.~(2009), the \Hb\ luminosity (Fang et al.~2009), 
the H$\gamma\lambda$434.0 nm
and H9$\lambda$383.5 nm luminosity (HH08), 
the NaI$\lambda$589.0 nm line luminosity (HH08)
and the \ion{He}{i}$\lambda$587.6 nm luminosity (Fang et al.~2009). 
An independent value of the mass accretion rate (log\Macc$-9.6\pm0.5$ \Msun/y) is
derived  from
the width of the \Ha\ line at 10\% of the peak ($\sim$340 km s$^{-1}$),
using the relation  derived by Natta et al.~(2004).

The NIR hydrogen recombination lines \Pab\ and \Pag\ 
have also been extensively used as proxy of the accretion luminosity
(e.g., Muzerolle et al. 1998); mass accretion rates from these
lines have been derived using the relationships
in Natta et al.~(2004)  and
Gatti et al.~(2008). All these values of \Macc\  are given in Table~\ref{parameters}.

The \ion{Ca}{ii} infrared triplet lines are 
attributed to accretion shocks very close to the stellar surface.
Mohanty et al.~(2005) showed that the \ion{Ca}{ii}$\lambda$866.2 is
a good proxy of the mass accretion rate in very low-mass stars; 
the value of  \Macc\ derived from their  
correlation for brown dwarfs and very low-mass objects is also
given in Table~\ref{parameters}, together with the
\Macc\ computed from 
the \ion{Ca}{ii} 854.2 nm line (HH08).

\begin{table}
\centering
\caption{Properties of selected lines. The EWs are measured relative to the 
observed local pseudo-continuum. The last two columns contain the flux and the 
derived $\dot M_{acc}$.}
\begin{tabular}{c c c c c}
\hline \hline
line & $\lambda$& pEW & Log Flux & log $\dot M_{acc}$\\
 & (nm) & (nm) & (erg/s/cm$^2$) & (M$_\odot$/y) \\
\hline
&&&&\\
\Hb\ & 486.1 & $-8.5\pm$ 0.7 & $-14.65$ & $-9.79\pm$0.44 \\
%\Ha\ & 656.3 & $-11.7\pm$ 0.7 & $-13.69$ &$-9.68\pm$0.53 \\	% 10%Ha
\Ha\ & 656.3 &$ -11.7\pm$ 0.7 & $-13.69$ &$-9.17\pm$0.56 \\	%LHa
H$\gamma$ & 434.0 &$-6.2\pm$ 0.2 & $-15.03$ &$-10.07\pm$0.42 \\
H9 & 383.5 &$-3.1\pm$ 0.6 & $-15.37$ &$-9.65\pm$0.37 \\
\ion{Ca}{ii} & 866.2 &$-0.19\pm$ 0.01 & $-14.93$ &$-9.65\pm$0.6 \\
\ion{Ca}{ii} & 854.2 &$-0.32\pm$ 0.01 & $-14.82$ &$-9.08\pm$1.0 \\
\ion{He}{i} & 587.6 &$-0.23\pm$ 0.02 & $-15.94$ &$-10.15\pm$0.88\\
NaI & 589.0 &$-0.18\pm$ 0.02 & $-16.17$ &$-10.24\pm$1.4\\
\Pab\ & 1281.8 &$-0.19\pm$ 0.01 & $-15.69$ &$-10.62\pm$ 1.41\\
\Pag\ & 1093.8 &$-0.13\pm$ 0.05 & $-15.63$ &$-10.43\pm$ 1.40\\

\hline\hline
\end{tabular}
\label{parameters}
\end{table}

The Balmer jump and Balmer continuum emission in the U-band are powerful diagnostics 
of accretion.  Plane-parallel slab models from Valenti et al. (1993) 
are used to calculate the bolometric correction, 
which is then applied to the excess continuum emission.
The excess emission is estimated by subtracting  
a photospheric template. 
We find log\Macc $\sim$ -9.94$\pm$0.4 \Myr following the procedure of HH08 
and adopting as photospheric template the class III object 
2MASS J05383858-0241558 (SO641 in Hernandez et al.~2007) 
observed close in time and airmass to J0538  
(spectral type M6.5, obtained from the X-shooter spectrum with the same procedure as J0538). 

\begin{figure}[!htb]
   \centering
%   \resizebox{\hsize}{!}{\includegraphics[]{comparison_NEW.epsi}}
 	\includegraphics[width=0.43\textwidth]{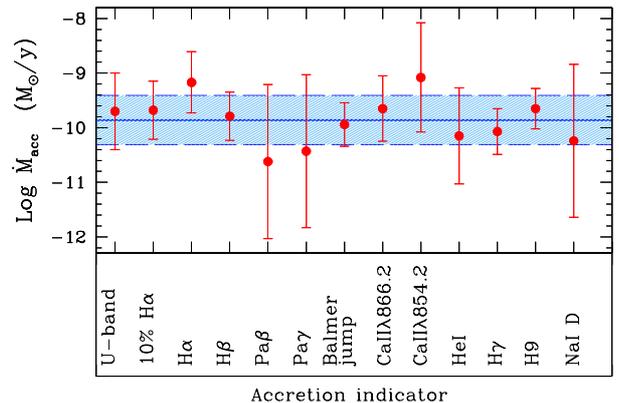} 
        \caption{\small Mass accretion rates determined from
different indicators, as labelled. 
The solid line shows the mean \Macc\ value, the shaded  region 
 the $\pm 1\sigma$ uncertainty. }
\label {comparison}
\end{figure}

Figure~\ref{comparison} shows the \Macc\ values  derived from the X-shooter
spectrum.
In most cases the uncertainties on \Macc\ are dominated by the
uncertainties on the adopted correlations with \Lacc\ (or, in some cases, \Macc).
Figure~\ref{comparison} also shows  
the mass accretion rate 
obtained by Rigliaco et al.~(2010) from the U band excess emission
(not simultaneous to the X-shooter spectrum),
recomputed 
for the spectroscopic spectral type derived in Sect.~3 (log\Macc $\sim -9.7\pm$0.7 \Myr). 

Even when accretion indicators are observed simultaneously,
the spread between different values of \Macc\ is quite large, 
of the order on $\pm 0.8$ dex. However, all values are
consistent within the large error bars, and the average \Macc\ has
an uncertainty of less than a factor of 3 (log\Macc $\sim -9.86\pm 0.45$ \Myr). 
On the other hand, the large scatter may reflect
a real spread in the values of the secondary indicators for 
fixed accretion luminosity, owing to, e.g., different stellar and/or disk
properties, different regimes of mass accretion, or differences in
wind/jet contributions.
When a significant number of X-shooter spectra of young objects that are distributed over a
large range of masses in different star-forming regions will become
available, it will be possible
to define and understand the relations between  \Lacc\ and the 
observable emission in lines and continuum much better.

\section {Mass loss rate}

The presence of forbidden lines such as [\ion{O}{i}]$\lambda$630.0 nm and
[\ion{S}{ii}]$\lambda\lambda$671.6,673.1 nm 
is an indication of jets or outflows from the system
(e.g., Hartigan et al.~1995; Whelan et al.~2009).
The most prominent optical forbidden line detected in the  J0538  spectrum
is the [\ion{O}{i}] line, while the [\ion{S}{ii}] lines
are not detected.
At the resolution of X-shooter,
the [\ion{O}{i}] line shows an asymmetric profile,
with stronger red-shifted than blue-shifted emission
(see Fig.~\ref{spectra}). 
From the [\ion{O}{i}] line luminosity we derive an average mass loss rate 
of  $\dot M_{wind} \sim 4.5 \times 10^{-12}$ \Myr,
following Hartigan et al.~(1995), including their assumptions that 
$V_{\perp} \sim 150 \, \rm{km\,\,s^{-1}}$ for the outflow
velocity, 
that the electron density is
$n_e$ $\sim 7.0 \times 10^{4}$ \cmq, (used because [\ion{S}{ii}] lines are not detected), 
and that the outflow fills
our 1-arcsec beam ($l_{\perp} \sim 5.4\times 10^{15}$ \, cm).

The value of $\dot{M}_{wind}$ is rather uncertain, given the
assumptions on $V_{\perp}$, $l_{\perp}$, and $n_e$.
However, the inferred ratio of the mass loss rate to the mass accretion rate ($\sim$0.03)
is typical of T Tauri stars
(Hartigan et al.~1995, White \& Hillenbrand~2004)
and agrees with the expectation 
of accretion-driven jets and winds (e.g., Shu et al.~1994; Pudritz et al.~2007).

\section {Physical condition of the emitting gas}

The many hydrogen emission lines detected in the J0538 spectrum can provide
information on the physical conditions of the emitting gas.

\begin{figure}[!htb]
   \centering
   %\resizebox{\hsize}{!}{\includegraphics[]{Hdelta_caseB_thick_thin_log_err.epsi}}
   \includegraphics[width=0.45\textwidth]{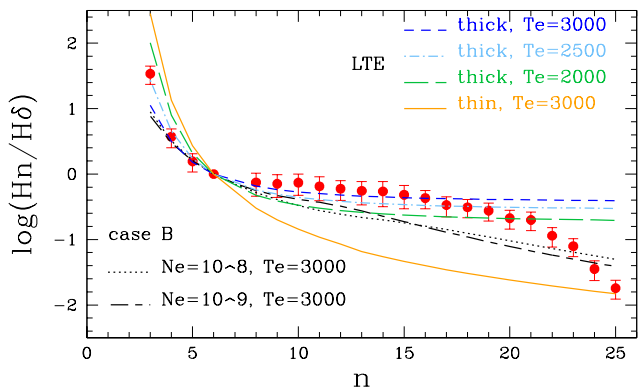}
        \caption{\small H$_n$/H$\delta$ line flux ratio versus the quantum number
$n$ of the upper level of the transition. The dots show the observed
values. Short-dashed, long-dashed and dot-dashed lines
plot the ratios for optically thick lines,  LTE level populations at
three different temperatures, as labelled. 
The solid line shows the results for LTE, optically thin lines. 
Predictions of Case B calculations for two different values of
the electron density and $T=3000$ K are shown by the
dotted and the short-long dashed line.  
%A color version of the figure is available in the online journal.
}
\label {decrement}
\end{figure}

Figure~\ref{decrement} shows the ratio of the Balmer line fluxes normalized
to H$\delta$, the strongest line with an almost symmetric profile.
For comparison, we computed the expected ratios
 for optically thick, LTE lines over a large range of temperatures, as well
as the Case B predictions, which assume that all lines are optically thin 
(Storey \& Hummer~1995) for different temperatures and
electron densities. We show in the figure the best-fitting curves.

The line ratios are well fitted by optically thick, LTE emission at relatively
low temperature (2000--3000 K). The highest Balmer lines ($n\ge$ 21)
are weaker than predicted when compared to \Hd, probably because they
become more and more 
optically thin. The Case B models  
do not provide an equally good fit.
The infrared lines are relatively weak, with \Brg\ not detected.
The observed values of \Pab/\Hg ($\sim 0.2$) and 
\Pag/\Hd ($\sim 0.4$), which refer to lines coming from the same upper
levels, are higher than expected  in the optically thin case,
but much lower than expected if all lines were optically thick and
coming from the same physical region. 
The observed ratio \Pab/\Pag  ($\sim 0.9$) is consistent with optically thick, LTE
gas at T$\sim 3000$ K. 
Our interpretation is that the
hydrogen recombination lines come from dense, relatively cold gas in thermal
equilibrium, with
different lines sampling regions of a different physical size.

The physical conditions of the emitting gas
inferred from the hydrogen spectrum are not consistent
with the predictions of the magnetospheric 
accretion models, in which the lines form at temperature of several thousand degrees 
(6000 $\leq$ T $\leq$ 12000 K, Muzerolle at al.~2001, 2005)
in the column of gas accreting onto the star. 
Gatti et al.~(2006) found that at low mass accretion rates the region emitting
the hydrogen lines is more likely to be the shocked photosphere, where the
infalling matter impacts onto the star.

\section {Conclusions}

We present the first observations of a young, accreting brown dwarf
in \sori\ (J053825.4-024241) obtained with the new 
spectrograph X-shooter, which provides a simultaneous, medium resolution, 
high-sensitivity spectrum over the entire wavelength range from $\sim$300 to $\sim$2500 nm.
The spectrum shows a large number of lines and excess continuum emission, which are 
typical signatures of the accretion-related phenomena that dominate the spectrum of young stars, as well as a number of absorption features that allow us to 
classify the object (M7$\pm 0.5$), estimate its mass (M$\sim 0.06$ \Msun), 
and confirm its youth.

We estimate the mass accretion rate from 12 different accretion indicators simultaneously observed with X-shooter.
The accretion rate determinations have a minimum-maximum spread 
of a factor of 30 for the different methods;
when using all values, we significantly 
reduce the uncertainty and obtain log\Macc= $-9.86\pm0.45$ \Myr,
similar to the
non-simultaneous determination from the
U-band excess emission (Rigliaco et al. 2010). 
J0538 is one of the most variable objects observed among   
very low-mass stars and brown dwarfs
in the \sori\ cluster (Caballero et al.~2006); the X-shooter spectrum proves
that the large discrepancies between \Macc\ values found by different techniques 
by various authors do not only depend on variability, 
but also on the uncertainties of the relations between the observed properties, 
such as line luminosities, and \Lacc.

The spectrum of J0538 shows emission in the [\ion{O}{i}]$\lambda$630.0 nm line, which we interpret as evidence
of mass ejection with log$\dot{M}_{wind}\sim -11.4$ \Myr. 
The ratio $\dot{M}_{wind}$/\Macc\ is $\sim 0.03$, which is low, but within the 
range observed in T Tauri stars (Hartigan et al. 1995).

The large number of hydrogen recombination lines in the X-shooter spectrum
allows us to study the physical conditions of the emitting gas.
We find that Balmer lines (up to $n\sim 21$) are optically thick and
likely produced in a cold (T$\sim 2000-3000$ K) dense region where LTE conditions hold.  
These conditions are not thought to be found in the accretion columns 
from the disk onto the central object, in which the gas  
is expected to be at several thousand degrees.
Gatti et al. (2006) found similar conditions in few brown dwarfs in Ophiuchus, accreting at low \Macc, and suggested that the lines (in that case \Pab\ and \Brg) were likely  formed in the
shocked photosphere of the central object. The results for J0538 indicate that
this may indeed be common, at least in very low-mass objects.
Bary et al.~(2008) obtained low-resolution infrared spectra of a number
of more massive T Tauri stars and compared the Paschen and Bracket decrement
to optically thick, LTE emission and to the predictions of Case B. They found
better agreement for Case B (i.e., for optically thin lines), for high density and temperatures similar to  the values we find in
J0538.  X-shooter will produce in the immediate future a large number of spectra of brown dwarfs and T Tauri stars with resolution and sensitivity much higher than those available to Bary et al. (2008), which will allow us to 
better constrain the accretion models.

\begin{acknowledgements}
We would like to dedicate this letter to the memory of Roberto Pallavicini, 
who prematurely passed away before seeing the X-shooter first light. 
\end{acknowledgements}
\bibliographystyle{aa}

\begin{thebibliography}{99}
%\bibitem{}Alexander,R.D., Clarke,C.J., Pringle,J.E., 2004, MNRAS, 354, 71
%\bibitem{}Abergel,A., Teyssier,D., et al.,~2003, A\&A, 410, 577
%\bibitem{}Zapatero-Osorio,M.R., B\'ejar,V.J.S., et al.,~2002, A\&A, 384, 937


\bibitem{}
Baraffe, I., Chabrier, G., Allard, F., et al.~1998, A\&A, 337, 403

%\bibitem{}
%Bary, J.S. \& Matt, S.P.~2007, IAUS, 243, 95

\bibitem{}
Bary, J.S., Matt, S.P., Skrutskie, M.F., et al.~2008, ApJ, 687, 376

%\bibitem{}
%Batalha, C.C., Stout-Batalha, N.M., Basri, G., Terra, M. 1996, ApJS, 103, 211

\bibitem{}
B\'ejar, V., Mart\'in, E., Zapatero-Osorio, M.R., et al.~2001,ApJ,556,830

\bibitem{}
B\'ejar, V., Zapatero Osorio, M.R., \& Rebolo, R. 2004, AN, 325, 705

\bibitem{}
Caballero,J., Martin,E., et al.~2006, A\&A, 445, 143

%\bibitem{}
%Caballero, J.A.~2008, A\&A, 478, 667

%\bibitem{}
%Calvet, N. \& Gullbring, E.~1998, ApJ, 509, 802

\bibitem{}
Calvet, N., Muzerolle, J., Brice\~no, C.~2004, AJ, 128, 1294

%\bibitem{}
%Cutri, R. M., Skrutskie, M.F., van Dyk, S., et al.~2003, 2MASS All Sky Catalog of Point Sources, The IRSA 2MASS All-Sky Point Source Catalog, NASA/IPAC Infrared Science Archive (Pasadena, CA: NASA/IPAC), http://irsa.ipac.caltech.edu/application/Gator

%\bibitem{}
%Dahm, S.E. 2008, AJ, 136, 521

\bibitem{}
Fang, M., van Boekel, R., Wang, W., et al. 2009, A\&A, 504, 461

%\bibitem{}
%Folha, D.F.M., \&\ Emerson, J.P.~2001, A\&A, 365, 90

%\bibitem{}
%Franciosini, E., Pallavicini, R., \& Sanz-Forcada, J.~2006, A\&A, 446, 501

\bibitem{}
Gatti, T., Testi, L., Natta, A., et al.~2006, A\&A, 460, 547

\bibitem{}
Gatti, T., Natta, A., Randich, S., et al.~2008, A\&A, 481, 423

\bibitem{}
Gullbring E., Hartmann L., Brice\~no C. \& Calvet N. 1998,ApJ,492,323

%\bibitem{}
%Gullbring, E., Calvet, N., Muzerolle, J., Hartmann, L.~2000, ApJ, 544, 927

\bibitem{}
Johnas, C., Guenther, E., Joergens, V., et al.~2007, A\&A, 475, 667

%\bibitem{}
%Hartigan, P., Kenyon, S.J., Hartmann, L., et al.~1991, ApJ, 382, 617

\bibitem{}
Hartigan, P., Edwards, S., Ghandour, L.~1995, ApJ, 452, 736

\bibitem{}
Herczeg, G. \&\ Hillenbrand, L. A.\ 2008, ApJ, 681, 594


\bibitem{}
Hern\'{a}ndez, J., Hartmann, L., Megeath, S., et al. 2007,ApJ,662,1067

%\bibitem{}
%Hummer, D.G. \& Storey,P.J.~1987, MNRAS, 224, 801

\bibitem{}
Kirkpatrick, J., Kelly, D., Rieke, G., et al.~1993, ApJ, 402, 643

\bibitem{}
Luhman, K.L., Liebert, J., \&\ Rieke, G.H.~1997, ApJ, 489, 165

\bibitem{}
Luhman, K.L. 1999, ApJ, 525, 466

%\bibitem{}
%Martin, S.C.~1996, ApJ, 470, 537

%\bibitem{}
%Mart\'in, E.L., Basri, G., Gallegos, J., et al.~1998, ApJ, 469, 706

\bibitem{}
Mohanty, S., Basri, G., Jayawardhana, R.~2005,Astron. Nachr.326,891

\bibitem{}
Muzerolle, J., Hartmann, L., Calvet, N.~1998, AJ, 116, 455

\bibitem{}
Muzerolle, J., Calvet, N., \&\ Hartmann, L. 2001, ApJ, 550, 994

\bibitem{}
Natta, A., Testi, L., Muzerolle, J., et al. 2004, A\&A, 424, 603

\bibitem{}
Muzerolle J., Luhman K.L., Brice\~no C., et al.~2005, ApJ 625, 906

\bibitem{}
Oliveira,  J., Jeffries, R., \& van Loon, J.~ 2004,MNRAS,347,1327

\bibitem{}
Palla, F., Randich, S., Pavlenko, Y.V., et al.,~2007, ApJ, 659, 41

\bibitem{}
Pudritz, R., Ouyed, R., Fendt, Ch., \&\ Brandenburg, A.~2007,PPV,277

\bibitem{}
Rigliaco, E., Natta, S., Randich, S., et al.~ 2010, A\&A in press

%\bibitem{}
%Sicilia-Aguilar, A., Hartmann, L., F\"ur\'esz, G., et al.~2006, AJ, 132, 2135

\bibitem{}
Shu, F.H., Nijita, J., Ostriker, E.~1994, ApJ, 419, 781

\bibitem{}
Storey,P.J. \& Hummer, D.G.~1995, MNRAS, 272, 41

\bibitem{}
Valenti, J., Basri, G., \&\ Johns, C.M.~1993, Aj, 106, 2024

\bibitem{}
Whelan, E.T., Ray, T.P., Podio, L., et al.~2009, ApJ, 706, 1054

\bibitem{}
White, R.J., \& Hillenbrand, L.A. 2004, ApJ, 616, 998



\end{thebibliography}

\end{document}